\documentclass[fleqn,12pt]{wlscirep}

\usepackage[utf8]{inputenc}
\usepackage[T1]{fontenc} 

\usepackage{graphicx}
\usepackage{miller}
\usepackage{siunitx} 
\usepackage{comment}
\usepackage{todonotes}
\newcommand{\degree}{$^{\circ}$}

\usepackage[yyyymmdd,hhmmss]{datetime}

\title{Standard-Based EBSD: Fingerprinting of Order and Orientation in Materials}

\author[1,a]{Aimo Winkelmann}
\author[2,b]{Grzegorz Cios}
\author[2,c]{Tomasz Tokarski}
\author[3,d]{Gert Nolze}
\author[4,e]{Ralf Hielscher}
\author[5]{Tomasz Kozie\l{}}

\affil[1]{Laser Zentrum Hannover e.V., Hollerithallee 8, 30419 Hannover, Germany}
\affil[2]{Academic Centre for Materials and Nanotechnology, AGH University of Science and Technology, al.\@ A. Mickiewicza 30, 30-059 Krakow, Poland}
\affil[3]{Federal Institute for Materials, Research and Testing (BAM), Unter den Eichen 87, 12205 Berlin, Germany}
\affil[4]{Technical University Chemnitz, Department of Mathematics, Reichenhainer Stra{\ss}e 39, 09126 Chemnitz, Germany}
\affil[5]{Faculty of Metals Engineering and Industrial Computer Science, AGH University of Science and Technology, al.\@ A. Mickiewicza 30, 30-059 Krakow, Poland}

\affil[a]{a.winkelmann@lzh.de}
\affil[b]{ciosu@agh.edu.pl}
\affil[c]{tokarski@agh.edu.pl}
\affil[d]{gert.nolze@bam.de}
\affil[e]{ralf.hielscher@mathematik.tu-chemnitz.de}

\begin{abstract}
Orientation determination does not necessarily require complete knowledge of the local atomic arrangement in a material. 
We present a method for microstructural phase discrimination and orientation analysis of phases for which there is only limited information available.
In this method, experimental Kikuchi diffraction patterns are utilized to generate self-consistent standards for use in the technique of Electron Backscatter Diffraction (EBSD). 
As an application example, we map the locally varying orientations in samples of icosahedral quasicrystals observed in a Ti40Zr40Ni20 alloy. 
\end{abstract}

\begin{document}

\flushbottom
\maketitle
\thispagestyle{empty}

\section*{Introduction}

The characterization of the microstructure of materials provides important information for the understanding of their properties.  
In this context, electron backscatter diffraction (EBSD) is a key tool for sub-micron-scale crystallographic analysis of materials in the scanning electron microscope (SEM).
EBSD delivers spatially resolved crystallographic information via measurement of backscattered Kikuchi diffraction (BKD) patterns that are formed by incoherent point sources of backscattered electrons within a crystal structure \cite{venables1973pm}.
Compared to other diffraction techniques, Kikuchi patterns have the distinct advantage that they can provide a rather extended, wide-angle view on potential point group symmetries of the phase that is probed locally by the incident electron beam.

If the crystal structure of the phases in the material is known, the properties of the corresponding Kikuchi patterns can be approximated using different theoretical models \cite{winkelmann2016iop}, and the local orientation can be determined by comparison of experimental Kikuchi pattern features to crystallographic predictions. 
In the analysis of complex microstructures, however, it can often be desirable to obtain orientational information even from phases for which we have only a limited knowledge about their actual crystal structure.

For high-precision microscopic orientation analyses by EBSD, pattern matching approaches apply a quantitative comparison of experimentally collected diffraction patterns with data simulated according to the theoretical crystal structure and additional physical parameters.
While advanced Kikuchi pattern simulations \cite{winkelmann2007um,callahan2013mm} can be extremely useful for a comparison to known structures, the application of such simulated patterns is limited by the knowledge of the crystal structure itself, but also by the ancillary parameters which determine the quantitative development of Kikuchi diffraction patterns from a specific material.
Even for known crystal structures, specific experimental effects can be very characteristic in a Kikuchi pattern but expensive to simulate at the same time. This includes, for example, features due to higher order Laue zone (HOLZ) rings \cite{michael2000um}, which can be very distinctive for a specific phase and electron energy.

In order to circumvent some of the aforementioned problems related to quantitative Kikuchi pattern simulations for arbitrary materials, we present an approach which generates a global diffraction standard from a number of experimentally measured Kikuchi diffraction patterns and subsequently uses this Kikuchi standard for a precise orientation determination in materials. 
A precise knowledge of the underlying atomic crystal structure is not needed for this standard-based EBSD orientation determination, as is illustrated by  earlier, non-automated, attempts on phase and orientation determination using atlases of Kikuchi pattern "fingerprints" in the absence of any sufficiently realistic simulations 
\cite{joy1982jap,dingley1995atlas}. 

The example application which we discuss in this paper concerns the local orientation determination in grains of quasicrystalline materials, which can provide additional information to understand the formation of such materials using different crystallization conditions or casting techniques 
\cite{tsai2015hbcg,steurer2018aca}. 
The locally resolved investigation of quasicrystals and related structures has been approached by EBSD in several previous studies \cite{naumovic2001prl,tanaka2016amat,bindi2016srep,kurtuldu2013amat,labib2019pm,singh2019pm}. 
Compared to these previous investigations, our standard-based EBSD approach reveals qualitatively new microstructural details even down to the sub-grain level, with a resolution that was beyond the reach of the existing studies. 
Apart from our specific application example, standard-based EBSD can be applied to a wide range of materials, providing an alternative method for phase discrimination and orientation analysis.

\section*{Experimental Kikuchi Diffraction Standards}

Kikuchi pattern features, such as Kikuchi bands and their intersections, directly correlate with the gnomonic projection of crystallographic features such as lattice planes \hkl(hkl) and lattice directions \hkl[uvw] on the planar phosphor screen. 
The lattice plane traces are related to the center of a Kikuchi band, and lattice directions correspond to the intersections of Kikuchi bands \cite{nolze2017jac}. 
Changes in these geometrically defined features of Kikuchi patterns can thus be linked to changes in the projection geometry, to lattice rotations, and to possible variations in the crystal lattice parameters.
\begin{figure}[tb]  %
	\includegraphics[width=0.9\textwidth]{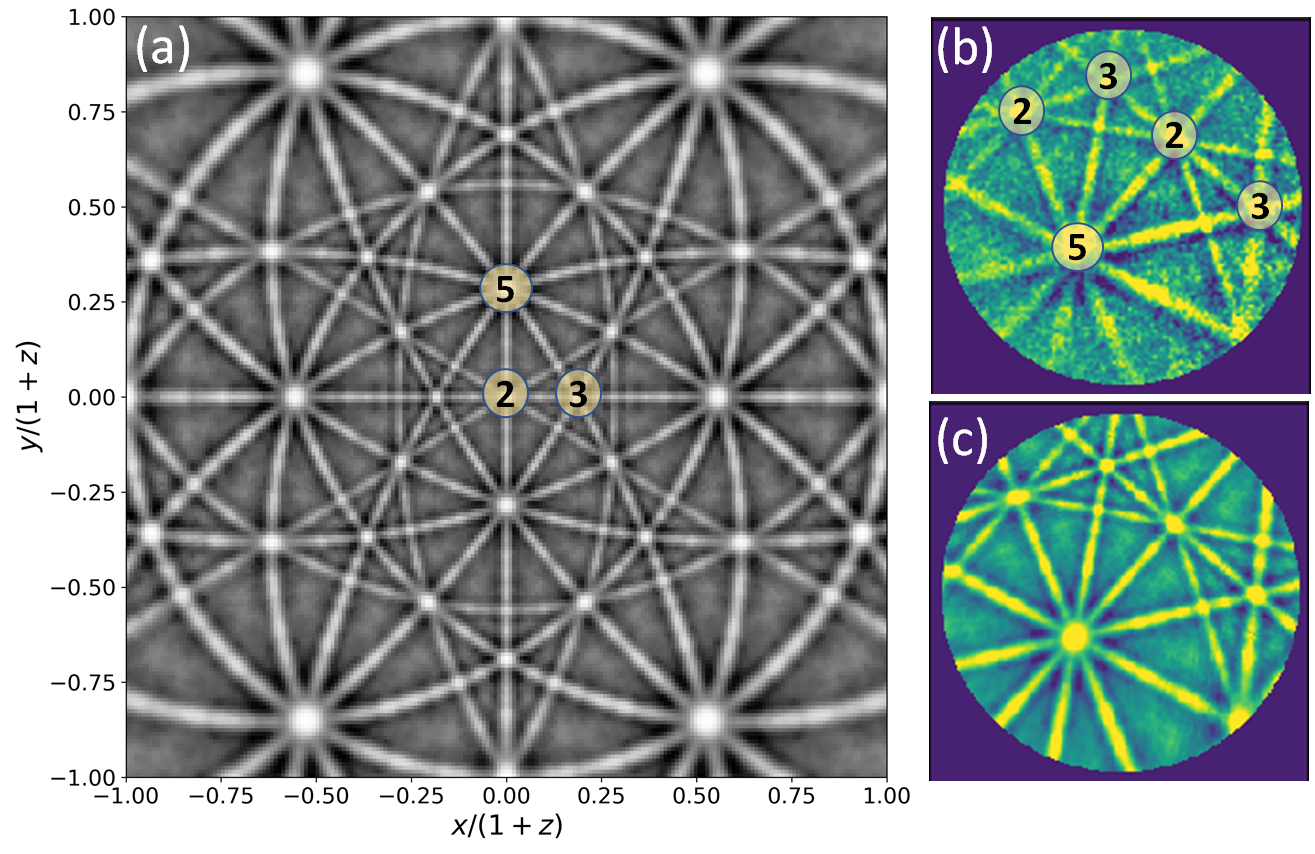}
	\caption{(a) Stereographic Kikuchi diffraction standard for quasicrystals formed in a Ti40Zr40Ni20 alloy, generated from the experimental pattern in (b) according to the icosahedral point group symmetry \cite{litvin1991aca}, (c) represents the best-fit reprojected pattern from the reference standard shown in (a).}
	\label{fig:sem}
\end{figure}
If we can identify a crystallographic reference frame from the observed directions in experimental Kikuchi patterns, we can thus also determine local crystallographic changes by comparison to purely experimental reference patterns.
This especially includes local rotations in the observed sample. 
Because the Kikuchi standard is energy-, phase- and material-specific, phase discrimination in general will also profit from standard-based EBSD. 

The generation of an experimental Kikuchi standard is illustrated in Fig.\,\ref{fig:sem} for an icosahedral   quasicrystal observed in a Ti40Zr40Ni20 (at$\%$) alloy. 
In panel (b), we show an experimental Kikuchi pattern with two-, three-, and five-fold rotation axes marked.  
With respect to the setting of the icosahedral point group as defined in \cite{litvin1991aca}, we reprojected the experimental data from (b) to the complete hemisphere as shown in the stereographic projection in (a). 
As is well known, a projective transformation in the plane is defined by a ($3\times3$)-matrix when we know the correspondences of at least 4 suitable point pairs \cite{hartley2003MVG} in the experimental coordinate system \cite{britton2016mc}.
In our case, we use the observed positions of the symmetry axes in (b) to map the experimental data to the corresponding reference directions and their symmetry-equivalent directions as given in \cite{litvin1991aca}. 
After this reprojection and symmetrization, the experimental data for the Kikuchi sphere \cite{day2008jmicr,day2009ebsd2,luehr2016nl,zhu2019arxiv} can serve as the global standard for the subsequent pattern matching \cite{chen2015mm,nolze2018amat}, bypassing theoretical simulations of the observed quasicrystal Kikuchi patterns.
As a side effect, the symmetrization of the experimental data also results in a reduction of noise, as can be seen by the reprojected Kikuchi pattern in Fig.\,\ref{fig:sem}\,(c). 
A comparison to the original pattern in Fig.\,\ref{fig:sem}\,(b) indicates that the low noise reprojection provides an optimized template for pattern matching. 

In comparison to the icosahedral Kikuchi standard shown in Fig. \ref{fig:sem}, the number of symmetry elements can be severely reduced for arbitrary other crystal structures.
In these cases, it will often be necessary to assemble the Kikuchi pattern standard from a sufficient number of calibrated experimental patterns. 
This is straightforward as long as the overlap between the measured patterns allows to identify the corresponding features on the Kikuchi sphere \cite{zhu2019arxiv}. 
Once the experimental Kikuchi standard has been obtained, however, standard-based EBSD pattern matching operates in a similar way with structures of any symmetry, enabling standard-based phase discrimination and orientation analysis.

\section*{Microtexture of a Quasicrystalline Material}

\begin{figure}[tb!]
\centering
    \includegraphics[trim={0 455 0 0}, clip, width=.33\textwidth]{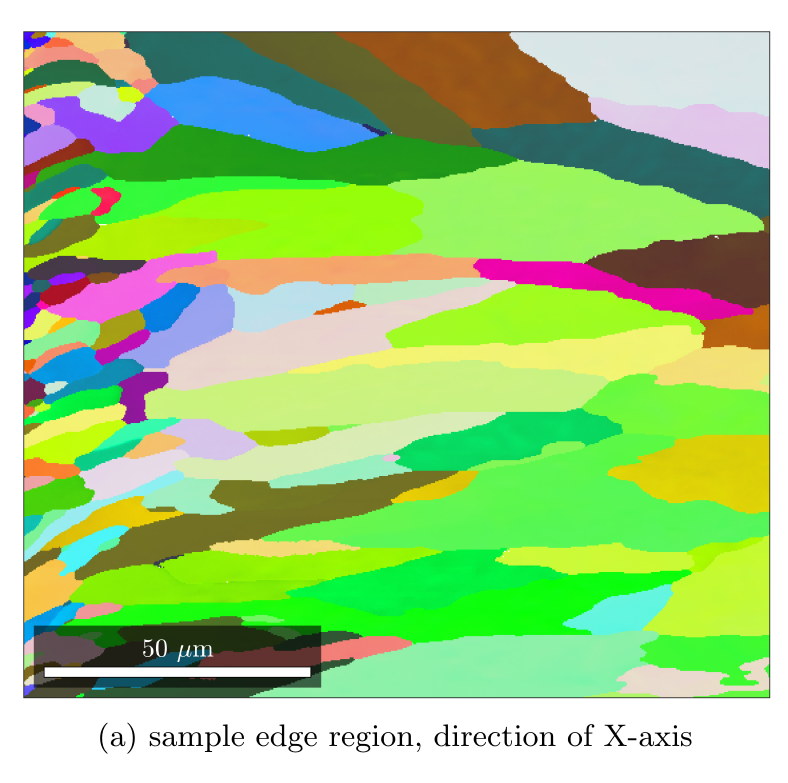}
    \hfill
    \includegraphics[trim={0 230 0 230}, clip, width=.33\textwidth]{ipf_xz.pdf}
    \hfill
    \includegraphics[trim={0 0 0 450}, clip, width=.32\textwidth]{ipf_xz.pdf}
    \caption{Orientation analysis of two different sample regions at the edge (a) and near the center (b) of a Ti40Zr40Ni20 sample. The colors correspond to the relative alignment of the sample X-axis (a) and Z-axis (b), in the reference system of point group 235 (c). The fundamental sector used is outlined using red lines in (c).}
	\label{fig:ipfy}
\end{figure}

As described in the Methods section below, the quasicrystalline Ti40Zr40Ni20 sample was produced by suction casting of the alloy melt into a cold copper mould, resulting in a cylindrical casting. 
The rapid solidification involves locally changing temperature gradients and differing cooling rates, which influences the competitive growth of quasicrystal grains. 
Based on what is known about the general properties of casting processes, we can expect the largest qualitative differences between the edge of the cast sample and its center \cite{glicksman2011solidification}. 
In our casting setup, the edge region of the sample is dominated by the high heat transfer through the copper mould walls in Fig.\,\ref{fig:ipfy}(a), while the sample center is controlled by a different temperature gradient leading to the larger and more equiaxed grains in Fig.\,\ref{fig:ipfy}(b).
In order to reveal the expected qualitative differences of the quasicrystal solidification near the sample edge compared to the center region, we investigated these two areas by EBSD using pattern matching to the experimental Kikuchi standard.
For an overview of applications of EBSD in the field of solidification, see \cite{boehm_courjault2009jmicr}.

The maps in Fig.\,\ref{fig:ipfy} show the results of the orientation determination using the icosahedral Kikuchi standard as determined in Fig.\,\ref{fig:sem}.
The color key in Fig.\,\ref{fig:ipfy}(c) is applied to color the map points according to the crystallographic direction which aligns with a given direction in the sample system, leading to so-called "inverse pole figure maps" ( "IPF-maps").
The measurement in Fig.\,\ref{fig:ipfy}(a) is near the sample edge, which is located at the left side of the map.
The temperature gradient was directed from the cylinder side walls towards the inside of the cylindrical sample, i.e. to the right of the measured area along the sample's X-axis. 
This is why, in Fig.\,\ref{fig:ipfy}(a), we use an IPF-X map to correlate properties of the inter-grain microtexture with the temperature gradient during solidification. 
As is typical for a cast microstructure, the grains in the nucleation zone at the left edge of the map in Fig.\,\ref{fig:ipfy}(a) are of small size, and randomly oriented as seen by the random color variation. 
After a short distance of competitive growth, grains with a fast growing direction parallel to the temperature gradient become dominant. 
The mainly greenish color in Fig.\,\ref{fig:ipfy}(a) indicates that a fast growing direction for this region is preferably parallel to one of the three-fold rotation axes, as can be seen in the color key in (c). 
The regions of grains with similar (in this case green) colors indicate a systematic, fiber-like orientation of these grains.
Different colors are observed for grains with their main axes not parallel to the temperature gradient. 

With a decreasing temperature gradient, the grains grow clearly larger, as can be seen in Fig.\,\ref{fig:ipfy}\,(b) for a sample region which is located in the center of the sample.
The coloring used in Fig.\,\ref{fig:ipfy}\,(b) refers to the normal direction parallel to the Z axis which was chosen to reflect the local temperature gradient. The change of the gradient direction from along the X-axis in Fig.\,\ref{fig:ipfy}\,(a) to along the Z-axis in Fig.\,\ref{fig:ipfy}\,(b) is consistent with the fact that the investigated sample has been taken from the lower part of the cast, where the heat transfer direction in the sample center can be additionally affected by the bottom crucible plate.
The red colors of the IPF-Z map shown in Fig.\,\ref{fig:ipfy}\,(b) now indicate a preferred growth along a two-fold rotation axis.

\begin{figure}[tb]  %
	\centering
	\includegraphics[trim={0 240 0 30}, clip, width=0.9\textwidth]{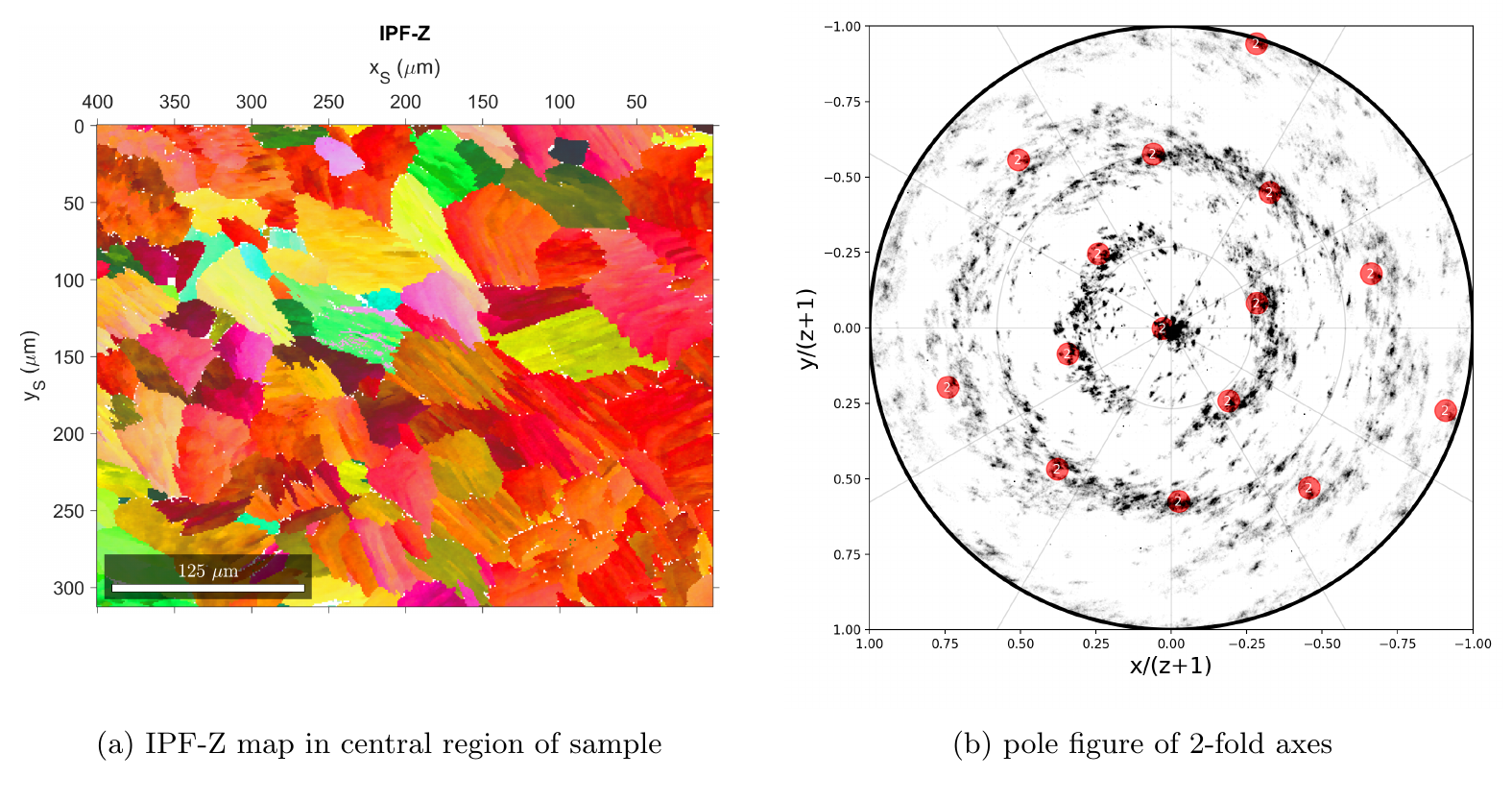}
	\caption{EBSD pole figure analysis from the center region of the Ti40Zr40Ni20 sample. The markers for the two-fold symmetry axes in (b) are shown for the mean orientation of the measurements in (a). The orientation changes in the mapped area can be approximated by systematic rotations around the two-fold axis pointing nearly parallel to the sample Z-axis, as indicated by the red colors of the IPF-Z map in (a), see also the color key in Fig.\,\ref{fig:ipfy}(c) .}
	\label{fig:map_polefig}
\end{figure}

Because the map in Fig.\,\ref{fig:ipfy}\,(b) does not allow a reliable prediction of a possible texture or preferred orientation due to the very low number of grains, in Fig.\,\ref{fig:map_polefig}\,(a) we show a larger map collected from a nearby position.
\begin{figure}[tb!]  %
	\hfill
	\includegraphics[width=0.85\textwidth]{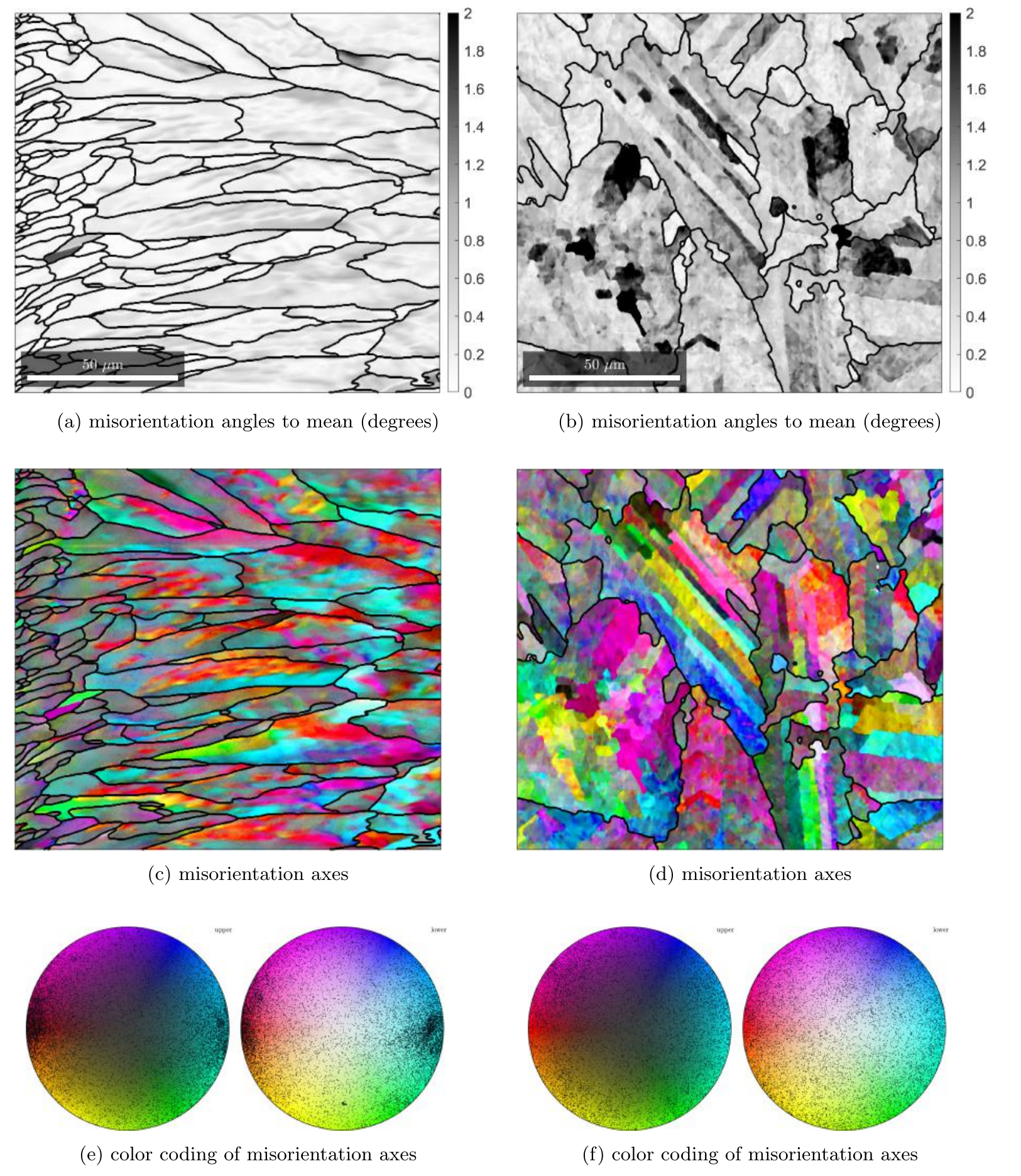}
	\hfill\mbox{}	
	\caption{Analysis of intragranular misorientations within the areas mapped in Fig.\,\ref{fig:ipfy} (left panels: edge region, right panels: central region). The misorientations are quantified by their rotation angle (top panels a,b) and the directions of the misorientation axes (middle, c,d) with the corresponding hemispherical color keys and axis distributions in the sample coordinate system (bottom, e,f). 
	The grain boundaries with a minimum misorientation angle of 5\degree{} are shown as the black lines.}
	\label{fig:miso}
\end{figure}
The grain boundaries which are visible in  Fig.\,\ref{fig:map_polefig}(a) suggest a dendritic growth which results in typically wave- or meandering-shaped boundaries. 
The pole figure in Fig.\,\ref{fig:map_polefig}\,(b) displays the distribution of all directions parallel to two-fold rotation axes. 
The fiber axis in the center of the pole figure means that the majority of grains in the map rotate around one of the two-fold axes which is perpendicular to the surface of the investigated sample.
We explicitly show by the filled circles the position of the two-fold axes for the mean orientation of all grains in Fig.\,\ref{fig:map_polefig}\,(a). 
As can be estimated from the pole figure, the inter-grain rotations cover a range of approximately $\pm$30 degrees around the mean.

In addition to the relative orientations of the grains in the maps shown in Fig.\,\ref{fig:ipfy} and \ref{fig:map_polefig}, we can also analyze the orientation variations \emph{inside} the individual grains. 
For example, in Fig.\,\ref{fig:ipfy}\,(b) as well as Fig.\,\ref{fig:map_polefig}\,(a) we observe a characteristic striation within the grains.
These variations can be described quantitatively by the orientation deviation relative to the mean orientation in the grain. 
In Fig.\,\ref{fig:miso}, we show the intra-grain misorientation analysis corresponding to the orientation maps in Fig.\,\ref{fig:ipfy}. 
The misorientations can be quantified by the value of the misorientation angles as seen in the top panels (a,b) of Fig.\,\ref{fig:miso} and by the directions of the misorientation axes relative to the mean orientation of the grain shown in the middle panels (c,d). The corresponding color keys for the upper and lower hemispheres in (e,f) are used to describe the 3D directions of the misorientation axes in the sample coordinate system as implemented in MTEX \cite{bachmann2010ssp}. A preferential alignment of misorientation axes will be seen by clustering of the mapped directions which are shown by the black dots. 

In the misorientation maps from the edge of the sample in the left part of  Fig.\,\ref{fig:miso}, we find continuous rotations within the grains, and no distinct subgrain boundaries are detectable, as is seen by the continuous color changes. 
However, the intragranular rotations indicate a clear correlation to the assumed fiber axis parallel to X because the misorientation axes cluster near the east-west axis in the hemispherical color key in Fig.\,\ref{fig:miso}(e).

We find a different situation in the central region of the sample, see the right panels (b,d,f) of Fig.\,\ref{fig:miso}. 
While the IPF-Z map in in Fig.\ref{fig:ipfy}\,(b) and the pole figure in Fig.\,\ref{fig:map_polefig} indicated a fiber-like texture relative to the sample Z-axis, the color jumps seen in Fig.\,\ref{fig:miso}(d) suggest subgrain boundaries which do not show an obvious correlation to the observed fiber texture. Consequently, the misorientation axis distribution does not display any pole clustering in Fig.\,\ref{fig:miso}(f). 
The lath-like subgrains in Fig.\,\ref{fig:miso}(d) display nearly constant colors which reflects a systematic misorientation of these blocks. The colors in the misorientation axis map in Fig.\,\ref{fig:miso}(d) also suggest that the small blocks with a diameter of the thickness of the lathes are likely to correspond to lathes a well, which are directed out-of-plane, however.
We expect that more insight into the peculiar intra-grain subtexture can be obtained by higher resolved EBSD measurements which could shed light on the possible role of minority phases and the formation of strain in the grains \cite{thi2006pre}. 
Specifically along the subgrain boundaries, Laves C14 and/or $\beta$-TiZr phases can occur, as these additional phases have been found previously by X-ray diffraction \cite{Qiang:03}. 
The local study of the distribution of the minority phases by EBSD in the future could also provide more general information on the rapid solidification process of the quasicrystalline phase with a related creation of local strains, because a rejection of chemical species is known to influence the attachment kinetics during quasicrystal growth \cite{thi2006pre,han2017srep}.

\section*{Discussion}

We have presented a standard-based EBSD technique which enables phase discrimination and texture determination based on experimental reference patterns, without the use of pattern simulations.
The necessary Kikuchi pattern standards can be generated by application of known symmetries to experimental Kikuchi patterns covering a fundamental sector of the full Kikuchi sphere.
Orientation analysis is then possible relative to the Kikuchi diffraction standard in a reference setting, which is preferably chosen in alignment with directions of symmetry axes. For the icosahedral quasicrystal analyzed here, these reference axes are chosen along the two-fold axes.  
Crystal orientations can be determined by template matching without explicit knowledge of the crystallographic indexing of lattice planes and directions.
A subsequent orientation refinement delivers a comparable orientation precision as pattern matching to simulated data \cite{nolze2018amat}.
Kikuchi standards can take into account the distinguishing experimental effects of the observed phases, which include for example, the formation of HOLZ rings \cite{michael2000um}. A proper inclusion of these effects in the Kikuchi standard will contribute to a higher orientation precision and a more sensitive phase discrimination in the pattern matching process.

The standard-based EBSD approach is best seen as an additional option which can be combined with other available methods for phase discrimination and orientation determination from Kikuchi patterns.
In this way, EBSD analyses from complex materials can benefit from the combination of approaches which are tailored to extract the optimum amount of information from the available experimental data. 
For example, initial orientations of very similar phases in many cases can be obtained using a conventional, fast, approach based on the Hough transform \cite{wright2000ebsd1}, while a detailed phase assignment could be based on additional verification by dynamical pattern simulations\cite{winkelmann2007um} or by comparison to available Kikuchi standards as presented here. 

The limits of the standard-based EBSD approach are set by effects which cannot be described in a simple reprojection of Kikuchi pattern templates from a single common reference standard. For example, the so-called excess-deficiency effects are related to the specific geometry of the incident beam and they can influence the assignment of polarity in non-centrosymmetric phases or lead to a systematic bias in the orientation determination. 
If patterns of a very similar orientation are compared, however, the experimental standard could also include the excess-deficiency effects and in this way lead to a higher reliability of polarity assignment in nanocrystalline structures, see e.g. \cite{naresh2019nl}. 
The available precision of a Kikuchi standard will be also limited by the generation of the pattern standard itself. This depends on the accuracy of the projection center description as well as on the precision of the relative alignment between several patterns which can be necessary to assemble the fundamental sector of the Kikuchi sphere. Also, pattern distortions other than those which are due to the gnomonic projection, need to be taken into account for a precise reprojection of experimental intensities to the spherical coordinate system of the Kikuchi standard.

We expect that the presented approach will be especially useful for texture analysis and phase discrimination in demanding materials like complex intermetallics, superlattice structures, as well as metastable materials, all of which can consist of ordered, partially disordered or deformed phases. 
Further possible applications include geological and biological materials, for which it can be difficult to obtain simulated patterns with a sufficiently detailed consideration of all experimental effects.

\vspace{0.5cm}
\mbox{Date:\, \date{\today \, \currenttime}}

\section*{Methods}

\subsection*{Materials and casting}

A mixture of high-purity elements with a composition of Ti40Zr40Ni20 has been arc melted in high vacuum evacuated argon atmosphere using an Arc Melter AM (Edmund B{\"u}hler GmbH). The solid sample was produced by suction casting into a copper mould of $\varnothing = 6$\,mm and a length of 55\,mm. Further information about the technique and apparatus can be found in \cite{Koziel:15}. 
The composition and casting technique are based on previous studies as published in \cite{Qiang:03}. 
The solidified  cylinder is polycrystalline, with grains of icosahedral quasicrystalline symmetry. 
The EBSD analysis was performed in a plane perpendicular to the longitudinal axis of the cast, which in the paper is referred to as the Z-axis.

\subsection*{EBSD Pattern Acquisition}

For the orientation mapping of the icosahedral quasicrystals, patterns with a resolution of $120\times 120$ pixel have been acquired using a Hikari camera (EDAX). A pattern averaging of 3$\times$ was used, resulting in an effective dwell time of 30\,ms. The detector was mounted on a Versa 3D (FEI) field-emission scanning electron microscope (FE-SEM) which was operated at 20\,kV acceleration voltage and 32\,nA beam current.
The raw EBSD patterns were stored for the subsequent data analysis steps.

\subsection*{Orientation Determination and Visualization}

For the orientation mapping, we determined an initial orientation description for each measurement point by a brute-force template matching approach similar to the one described in \cite{chen2015mm}.
We used a predefined set of diffraction pattern templates which covered the orientations of the point group 235 with an orientation resolution of 2.5\degree.  
The local projection centers have been determined for each map position according to a projective model of the SEM beam scan on the tilted sample plane \cite{britton2016mc}. 
For the final, high-precision orientation refinement, we applied nonlinear optimization using real-time reprojection \cite{nolze2018amat} of diffraction data and with starting values determined by the intial brute-force pattern matching step.
In the pattern matching, the best fit is selected by the largest value of the normalized cross-correlation coefficient $r$ \cite{tao2005mm} between the background-processed \cite{winkelmann2019prb} experimental pattern  and the template derived from the Kikuchi standard.
The presentation of the orientation results is based on color keys which reflect the rotational symmetry of the icosahedral quasi crystals in point group 235 \cite{litvin1991aca}, as implemented in MTEX 5.2 \cite{bachmann2010ssp}.

\newpage


%

\section*{Acknowledgements}

We would like to thank M. Buchheim and R. Saliwan-Neumann (BAM Berlin) for sample preparation and data collection. The project is partially financed by the Polish National Agency for Academic Exchange grant no. PPI/APM/2018/1/00049/U/001.

\section*{Author contributions statement}

A.W. conceived the data analysis approach, T.K., G.C., T.T., and G.N. conducted the experiments and analysed the results.  R.H. contributed the orientation visualization for quasicrystal symmetries. All authors reviewed the manuscript.

\section*{Additional information}

\textbf{Competing interests} The authors declare no competing interests. 

\end{document}